\documentclass[12pt]{article}

\usepackage[pdftex]{graphicx}

\usepackage{scicite}
\usepackage{times}
\usepackage{bm}
\usepackage{amssymb}
\usepackage{graphicx}
\usepackage{color}
\usepackage{caption}
\newenvironment{sciabstract}{%
\begin{quote} \bf}
{\end{quote}}

\title{MD-GAN with multi-particle input:\\
the machine learning of long-time molecular behavior from short-time MD data} 

\author
{Ryo Kawada,$^{1}$ Katsuhiro Endo,$^{1}$ Daisuke Yuhara,$^{2}$ Kenji Yasuoka$^{1}$\\ 
\\
\normalsize{$^{1}$Department of Mechanical Engineering, Keio University,}\\
\normalsize{3-14-1 Hiyoshi, Kohoku-ku, Yokohama, Kanagawa, 223-8522, Japan}\\
\normalsize{$^{2}$Science \& Innovation Center, Mitsubishi Chemical Corporation, Yokohama, 227-8502, Japan}\\
\\
}

\date{}

\begin{document}

\baselineskip24pt

\maketitle

\begin{sciabstract}
MD-GAN is a machine learning-based method that can evolve part of the system at any time step, accelerating the generation of molecular dynamics data. For the accurate prediction of MD-GAN, sufficient information on the dynamics of a part of the system should be included with the training data. Therefore, the selection of the part of the system is important for efficient learning. In a previous study, only one particle (or vector) of each molecule was extracted as part of the system. Therefore, we investigated the effectiveness of adding information from other particles to the learning process. In the experiment of the polyethylene system, when the dynamics of three particles of each molecule were used, the diffusion was successfully predicted using one-third of the time length of the training data, compared to the single-particle input. Surprisingly, the unobserved transition of diffusion in the training data was also predicted using this method.
\end{sciabstract}

\clearpage

\section*{Introduction}
Molecular dynamics (MD) simulation is a method of investigating the behavior of molecules by numerically solving their classical equations of motion, which is useful for analyzing a variety of structural and dynamic properties and even mechanisms of phenomena. MD simulations have been applied to a wide spectrum of materials such as biomolecules\cite{protein}, polymers\cite{polymer}, carbon nanotubes\cite{cnt}, and methane hydrate\cite{methanehyd}. In research using MD simulations, the huge computational cost of large-scale and long-time simulations is a long-standing problem that must be addressed to expand possible applications. Massively parallel computation by dividing space\cite{parallel}, coarse-grained simulation\cite{Kremer}, and multiple time-step method for efficient time evolution\cite{multistep} are valid methods for reducing the computational cost of MD simulations. However, including these techniques does not reduce the cost greatly, particularly in long-time simulations. One possible solution to this problem, which has recently attracted attention, is the use of machine learning. In this study, we have focused on machine learning techniques for MD simulations.

In the past few years, many studies have reported the use of machine learning for the analysis and calculation of MD trajectories\cite{Noea}. The construction of a force field model with machine learning\cite{Behler,Bartok,Chmielaa,Behlera,Li,Chmiela,Han}, the estimation of the free energy surface\cite{Stecher,Schneider}, and the detection of specific molecular behavior\cite{nanoscale} are useful methods to improve the accuracy of MD simulations and the efficiency of MD trajectory analysis. A major issue increasing the computation cost of MD simulations is the presence of a complex energy plane geometry that has multiple high-energy barriers. To solve this problem, enhanced sampling methods have been proposed, and machine learning techniques have been applied to further improve the sampling efficiency\cite{Ribeiro,Mones,Valsson,Bonati}. Moreover, the construction of Markov state models from MD trajectory data is useful for describing long-term behavior of dynamical systems. Methods to estimate Markov state models using a machine-learning approach have also been developed recently\cite{Mardt,Chen,Wu}. Furthermore, methods based on graphical models (dynamic ones)\cite{Olsson} and flow-based models (Boltzmann generator)\cite{boltzmanngen} have been introduced, and the sampling of unobserved states has been achieved. These proposed methods are notable in terms of improvements in sampling efficiency; however, they require a large amount of MD trajectory data for training. In contrast, a proposed generative model for MD trajectories (MD-GAN)\cite{endo} has achieved direct and efficient prediction of long-term molecular behavior from short-time MD trajectory data. In this study, we have focused on MD-GAN.

MD-GAN is based on a machine learning model called generative adversarial network (GAN)\cite{gan}, which consists of two networks: a generator and a discriminator. Short-trajectory data are prepared by the MD simulation as training data. In the training phase, the generator is trained to generate a time series of the molecular behavior, whereas the discriminator is trained to evaluate whether the generated data are real or fake. After training, the generator can be used repeatedly to produce long-time series data. To reduce the computational cost of generating long-time series data compared to direct MD simulations, the input training data are preprocessed. MD simulations reproduce the behavior of ``all particles'' on ``short-time steps (fs order)''; however, all information is not usually required to calculate the physical properties. For example, when analyzing the diffusion of entangled polymer molecules, the ps--ns order behavior of the central atom of each molecule is sufficient. Therefore, in the preprocessing of the input data for MD-GAN, only the particle information of interest is extracted (feature extraction), and the information on short timescales is thinned out (step skip). The behavior of the target particle in long-time steps ($>$ fs order) is learned and generated by MD-GAN; therefore, it is much faster than direct MD simulations. Figure 1A shows the flow of generating long time-series data using MD-GAN. With these mechanisms, a speed-up of 19 times is achieved in the prediction of polyethylene (PE) diffusion compared with direct MD simulations\cite{endo}.

Although MD-GAN is efficient for the prediction of physical properties, the time required for the training data is still problematic. For PE diffusion, a relatively long MD dataset is used as the training data. It is known that the diffusion of entangled polymer molecules shows a transition from abnormal to normal diffusion as the time scale increases. This is due to the slippage of molecular chains occurring over a longer time scale\cite{takahashiamd}. Information about this transition must be included in the training data to accurately predict long-time-scale diffusion. In the PE system of a previous study, the transition of diffusion occurs at 1 ns or longer time scales. Hence, nanosecond order simulations are required to be performed for the generation of training data and are usually difficult to execute for polymer systems.

In this study, we aimed to find a way to make MD-GAN prediction possible even with shorter MD data. In a previous study, only one target particle’s behavior of each molecule was extracted for the preprocessing of input data. Hence, we assumed that one of the possible solutions is the addition of the information of other particles to the input data. In the entangled polymer diffusion case, we considered that when the dynamics of some particles in the same molecule are added to the input data, sufficient information on polymer chain dynamics can be obtained with shorter MD data (see Figure 1B). We improved the architecture of the MD-GAN for multi-particle input and examined whether the time length of the training data could be reduced. Performing the experiment on an entangled PE chain system, the prediction of diffusion from shorter-time data compared with a previous study was achieved using a three-particle input. Surprisingly, the unobserved transition of diffusion was successfully predicted, even though the time scale of the training data was limited to the abnormal diffusion region.

\section*{Methods}
\subsection*{Architecture of MD-GAN}
The flow of prediction of molecular behavior using MD-GAN is shown in Figure 1A. For preprocessing of the training data (short-time MD data), the $M$ steps of the trajectory of one particle (or the vector between two particles) are extracted from the prepared MD data (feature extraction), skipping several time steps (step skip). The time evolution of this extracted system (subsystem) is achieved by MD-GAN. The computational cost of this time evolution is lower than that of the direct MD simulation. The goal of the MD-GAN training is to sample
\begin{eqnarray}
    \bm{Y}_{(k+1)*M:(k+2)*M-1}\sim  p\left(\bm{Y}_{(k+1)*M:(k+2)*M-1}|\bm{Y}_{k*M:(k+1)*M-1}\right)\label{eq:intro1}
\end{eqnarray}
where $\bm{Y}_{a:b}$ is the time series of a subsystem from time step $a$ to time step $b$. $k$ is an integer greater than or equal to 0. Here, we make the following three assumptions:
\begin{itemize}
    \item The next $M$ steps of the subsystem depend only on the previous $M$ steps of the subsystem. (Markov property)
    \item Stochastic evolution of the subsystem is time invariant. This implies $p\left(\bm{Y}_{(k+1)*M:(k+2)*M-1}|\right.$\\$\left.\bm{Y}_{k*M:(k+1)*M-1}\right)$ is independent of $k$.
    \item The probability distribution of $M$ steps of the subsystem is stationary. This implies that the probability distribution represented by $p\left(\bm{Y}_{k*M:(k+1)*M-1}\right)$ is independent of $k$.
\end{itemize}
With these assumptions, Eq. \ref{eq:intro1} becomes simple,
\begin{eqnarray}
    \bm{Y}_{M:2M-1}\sim  p\left(\bm{Y}_{M:2M-1}|\bm{Y}_{0:M-1}\right)\label{eq:intro2}
\end{eqnarray}
This stochastic evolution of the subsystem is achieved by training the MD-GAN with short MD data in $2M$ steps. After the training, a long time series of subsystems is achieved upon iterative usage of the generator. However, we encounter an exposure bias if we use the generator many times. In multistep generation, the previous output by the generator is used as the real input, and this includes some small bias caused by the incompleteness of the generator. The accumulation of this small bias by multistep generation is non-negligible, and many studies have been conducted to reduce it\cite{exposurebias1,exposurebias2,exposurebias3}. MD-GAN succeeds in reducing the exposure bias by introducing latent variables and distribution stabilization mechanisms. In general, most real-world data can be considered as lying along a low-dimensional manifold in a high-dimensional space\cite{latent1,latent2,latent3}. Based on this idea, the evolution of the extracted trajectory of the subsystem is also represented as the evolution of dense latent variables, embedded in a low-dimensional space. Figure 2A shows the effects of these latent variables. By performing time evolution in such a low-dimensional dense space, deviations from the manifold can be mitigated. Figure 2B shows the architecture of the MD-GAN. Generator $G_z$ evolves the latent variables in time, and generator $G_Y$ transforms the latent variables into trajectories of the subsystems. The discriminator calculates the Wasserstein distance\cite{wgan} between the two generated M-step trajectories, and the short training data input to the MD-GAN. The initial latent variable $z_1$ is sampled from the W-dimensional uniform distribution $U(R^w)$. $G_z$ generates further $z_i$ by using the previous $z_(i-1)$ and a random value sampled from $U(R^w)$. We then apply $G_z$ to $z_1$ and repeatedly generate $z_2$, $z_3$, and so forth. If the distribution of the latent variables is not stationary in this way, the third assumption of the MD-GAN would not be satisfied. Therefore, $G_z$ is applied repeatedly until the distribution of latent variables becomes stationary (seven times or more is sufficient).

MD-GAN is based on WGAN-GP, which uses the Wasserstein loss formulation and a gradient norm penalty to achieve Lipschitz continuity\cite{wgan-gp}. The internal structures of the generators and discriminators are described in the Supplementary Materials (see Section S1). In WGAN-GP, the loss function is defined as
\begin{eqnarray}
    L={\mathbb{E}}_{\tilde{\bm x}\sim \mathbb{P}_{g}}\left[D(\tilde{\bm x})\right]
    -{\mathbb{E}}_{\bm x\sim \mathbb{P}_r}\left[D(\bm x)\right]
    +\lambda{\mathbb{E}}_{\hat{\bm{x}}\sim {\mathbb{P}}_{\hat{\bm{x}}}}
    \left[\left(\|\nabla_{\hat{\bm{x}}}D\left(\hat{\bm{x}}\right)\|_2-1\right)^2\right]
    \label{eq:wgan-gp}
\end{eqnarray}
where $\mathbb{P}_{g}$ is the generator distribution, $\mathbb{P}_{r}$ is the data distribution, and $D$ is the discriminator. $\mathbb{P}_{\hat{\bm{x}}}$ is defined as the distribution of ${\hat{\bm{x}}}$ uniformly sampled along a straight line between a pair of points sampled from the data distribution $\mathbb{P}_{g}$ and the generator distribution $\mathbb{P}_{r}$. $\lambda$ is a coefficient. The discriminator was optimized to maximize $L$, whereas $G_z$ and $G_Y$ were optimized to minimize $L$. 

\subsection*{Improvement of MD-GAN}
As mentioned in the Introduction, we assumed that the multi-particle input enhances the learning of long-time scale molecular behavior from short-time data. Thus, the architecture of MD-GAN was improved for multiparticle input. In this study, the validity of multi-particle input was examined by predicting PE diffusion.

\section*{Results}
\subsection*{Prediction of the PE diffusion by single particle input}
In this study, PE diffusion was predicted using the MD-GAN. The results of the MD simulation performed by Takahashi et al.\cite{takahashiamd,takahashi} were used for the training and evaluation of the prediction accuracy of the MD-GAN. The system consisted of 300 PEs with 1405 g/mol of 100 united-atom particles in a row, and the simulation data were available for 500 ns. The details of the MD simulations are described in the Supplementary Materials (Section S2). The diffusion coefficient is calculated as the slope of the mean square displacement (MSD) of the chain center:
\begin{eqnarray}
\label{eq:msd}
  	MSD(t)=\langle\left[\bm{r_{c.c.}}(t)-\bm{r_{c.c.}}(0)\right]^2\rangle
\end{eqnarray}
where $\bm{r_{c.c.}}(t)$ is the coordinate of the center of chain at time $t$. The diffusion coefficient at the center of the chain is:
\begin{eqnarray}
\label{eq:dc}
  	D=\lim_{t \to \infty}\frac{MSD(t)}{6t}
\end{eqnarray}
The diffusion behavior of the chain-center anomaly varies with time owing to chain entanglement\cite{takahashi}. In reptation theory\cite{reptation}, chain entanglement is described as a tube constant, and the polymer chain can only move along the major axis like a snake (reptation motion). The MSD calculated by the MD simulation is shown in Figure 3A. When $MSD(t)\propto t^n$, $n$ can be calculated from the slope of the MSD on both logarithmic graphs using the least squares method. The relationship between $t$ and $n$ is shown in Figure 3B, in which the transition of $n$ from 1/2(abnormal) to 1(normal) can be confirmed.

First, we confirmed the time length of the short-time MD data required for the single-particle input of the MD-GAN. The coordinates of each PE chain center were used as input for the MD-GAN. $M$ for Eq. \ref{eq:intro1} was set to 64. The number of iterations was set to 300,000, and a trajectory was generated for every 10,000 iterations. The length of the short-term MD data input to the MD-GAN can be adjusted by adjusting the number of step skips. Statistical fluctuations in the generated results were observed even when the same hyperparameter values were applied. This was because random numbers were used in the MD-GAN. Therefore, the MD-GAN was run five times under each condition, and its accuracy was statistically evaluated.

The prediction accuracy of MD-GAN was evaluated using the mean absolute percentage error(MAPE) in the transition region of the MSD. The MAPE at time t is given by
\begin{eqnarray}
    f(t)=\left|\frac{MSD_{\rm{gen}}(t)-MSD_{\rm{md}}(t)}{MSD_{\rm{md}}(t)}\right|\times 100
  	\label{eq:mape1}
\end{eqnarray}
where $MSD_{\rm{md}}(t)$ is the MSD at time $t$ obtained from the long-term MD data and $MSD_{\rm{gen}}(t)$ is the MSD at time $t$ obtained from the trajectory generated by the MD-GAN. Integrating the transition region on a logarithmic scale and taking the average to obtain
\begin{eqnarray}
    err := \int_{s_{1}}^{s_{2}} f(t) ds / \int_{s_{1}}^{s_{2}} ds
  	\label{eq:mape2}
\end{eqnarray}
where $s=\log_{10}t$, $s_1$ corresponds to the time when $n=1/2$ diffusion ends, transition to $n=1$ diffusion begins, and $s_2$ corresponds to the time when the transition ends. From Figure 3B, $s_1$ and $s_2$ in Eq. \ref{eq:mape2} were set to 1 ns and 30 ns, respectively. The methods for discretizing Eq. \ref{eq:mape2} are described in the Supplementary Materials (Section S3).

Various time lengths of the MD data were used as training data, and $err$ (Eq. \ref{eq:mape2}) was calculated for each case. The dimension of the latent variable ($n_z$) was set to 4 in all cases. When the time length of the input MD data was 2816 ps, the average $err$ (average of the smallest $err$ among the five generated trajectories) was 1.7 \%. The MSD of $err = 1.6$ \% case is shown in Figure 4A. The MSD shows a good agreement between the direct MD result and the prediction result by MD-GAN. On the other hand, when the time length of the input MD data was 1024 ps, the average $err$ was 15.3 \%. Figure 4B shows the MSD of $err = 14.7$ \% case. The MSD at long time scale shows a deviation from the MD result. It means that the information of the transition of diffusion (as shown in Figure 3) was not enough in the 1024 ps data of chain center motion.

\subsection*{Improvement by multi-particle input}
To investigate the validity of the multi-particle input, the 15th and 85th particles of the chain were added to the input. A snapshot of the PE chain system and the selected particles is shown in Figure 5. In the three-particle input, the relative coordinates from the chain center were used to describe the positions of the particles. By adding the relative coordinates of these particles, the MD-GAN is expected to learn information corresponding to the reptation motion. The length of the MD data was set to 1024 ps, which is the same as when the prediction by the single-particle input failed. It is noted that the MSD was calculated only from the behavior of the chain-centered particles, although the behavior of the three particles was trained and predicted.

The average $err$ of the five runs was 54.2 \%, which was much larger than that of the single-particle case. The reason for this deterioration was considered to be the four-dimensional latent variable, which was too small to adequately retain information about the dynamics of the three particles. Thus, we investigated the appropriate value of $n_z$ for the three-particle input. The average $err$ for various $n_z$ in the single- and three-particle inputs is shown in Figure 6. Although the average $err$ of the single-particle input was not changed for each $n_z$, that of the three-particle input dramatically decreased to 10 \% or below when $n_z$ was 10 or greater. When $n_z$ was 64, the average $err$ was 2.1 \%, and the MSD with an $err$ of 2.0 \% is shown in Figure 7. The deviation of the MSD over a long-time scale, as shown in Figure 4B, was successfully fixed by the three-particle input. Although 2816 ps of MD data was required for accurate prediction with single-particle input, the time length of the input data was reduced to approximately one-third.

The selection of three particles for the input is also important. We selected both sides of the chain center instead of the 15th and 85th particles of the chain. As a result, the average $err$ of the five runs was 22.1 \%, which was worse than that of the single-particle input. This implies that an increase in the number of particles is insufficient, and the selection of particles is important to achieve the motion of the polymer chain.

\section*{Discussion}
In this study, by inputting the information of multiple particle dynamics into the MD-GAN, the time length of the MD data, to be prepared in advance, was reduced compared to the case of inputting that of a single particle. In the case study of the PE system, when the dynamics of the three particles were inputted, the diffusion of the chain center was successfully predicted by approximately one-third of the length of the MD data compared with the single-particle input. The unobserved transition of diffusion was predicted, even though the time scale of the training data for the three-particle input was limited to the abnormal diffusion region. In addition, the selection of the three particles is important because if particles very close to the chain center particle are selected, a decrease in the prediction accuracy will occur. This indicates that multiple particles must be adequately selected to express polymer chain motion.

It should be noted that although the required time length of the training MD data was reduced by the multi-particle input, the total computational cost actually increased. The increase in the computational cost was mainly caused by the increase in the dimension of the latent variable, and the training took a significant amount of time. However, the computational cost of training can be reduced using various techniques, such as an increase in batch size\cite{Smith} and large-scale parallel computing. Therefore, the reduction in computational cost for the generation of training MD data achieved in this study is useful for accelerating MD-GAN. 

In future work, we will apply MD-GAN with multiparticle input to various molecular systems and find an appropriate way to select multiple particles to predict the physical properties of each system.

\section*{Supplementary Materials}
Section S1. Detailed architecture of MD-GAN\\
Section S2. Details of MD simulations\\
Section S3. How to calculate $err$\\
Fig. S1. Detailed architecture of MD-GAN


\bibliography{kwdmain}
\bibliographystyle{ScienceAdvances}

\section*{Acknowledgments}
We are grateful to Dr. Taku Ozawa, Mr. Yoshitsugu Kakemoto, and Mr. Kousuke Ohata (JSOL) for the many stimulating discussions that we have had. This work was supported in part by the Ministry of Education, Culture, Sports, Science, and Technology (MEXT) as Research and Development of Next-Generation Fields.

\clearpage
\begin{figure*}[!ht]
\begin{center}
\includegraphics[width= 15cm]{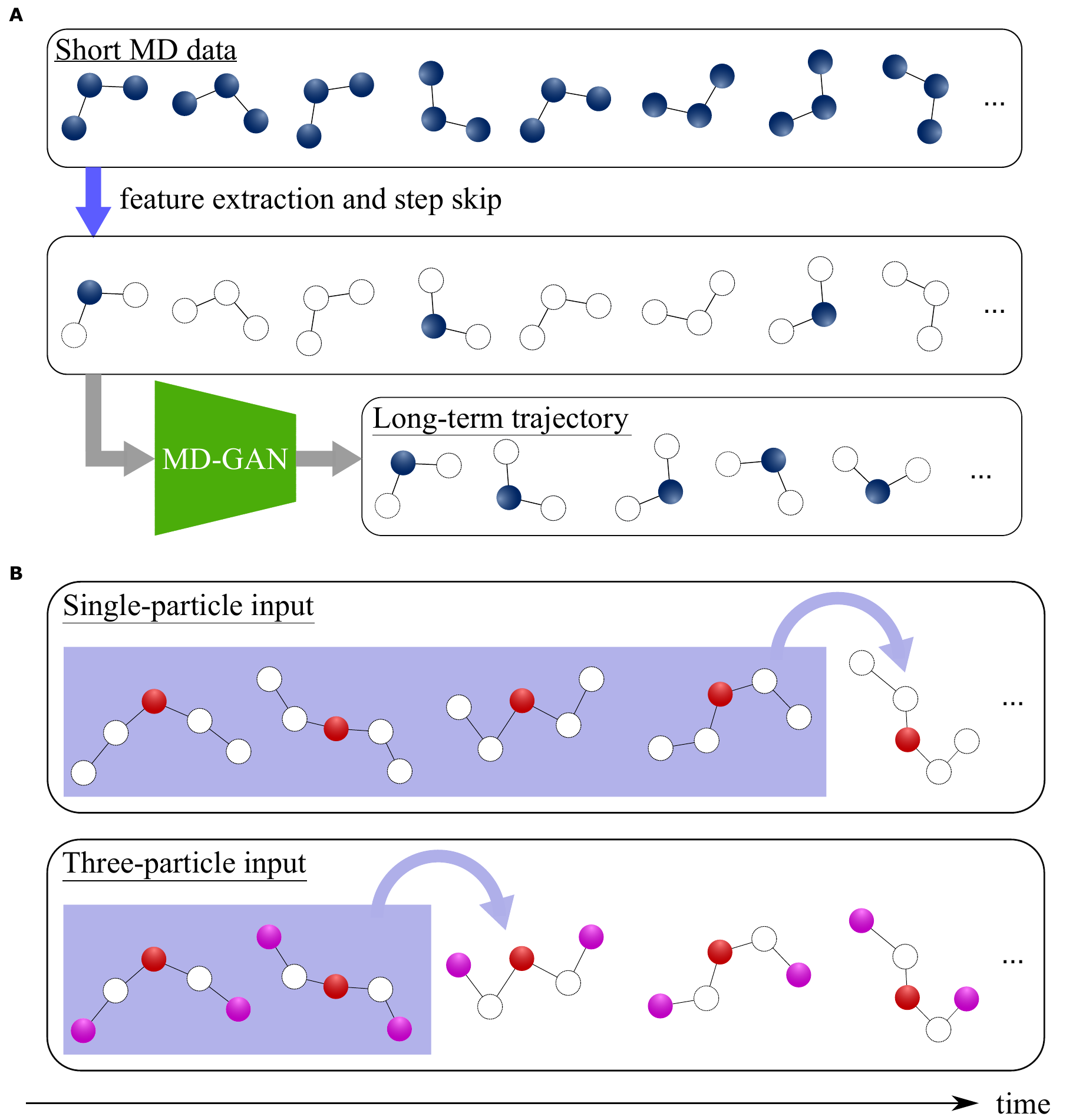}
\caption{
{\bf Schematic view of the learning and the prediction procedure of MD-GAN.} 
{\bf (A)} Flow of learning and prediction using MD-GAN. The information about single particle dynamics extracted by feature extraction and step skip is learned by MD-GAN, and the long-term trajectory is generated. 
{\bf (B)} Concept of multi-particle input. The single particle input considers the motion of only the red particle in one molecule, while the three-particle input considers the motion of the red particle with the two pink particles. The blue shaded ranges indicate the time length of input data. We considered that sufficient information for molecular dynamics could be achieved from shorter MD data if multi-particle input was applied.
}
\label{fig1}
\end{center}
\end{figure*}

\begin{figure*}[t!]
\begin{center}
\includegraphics[width= 15cm]{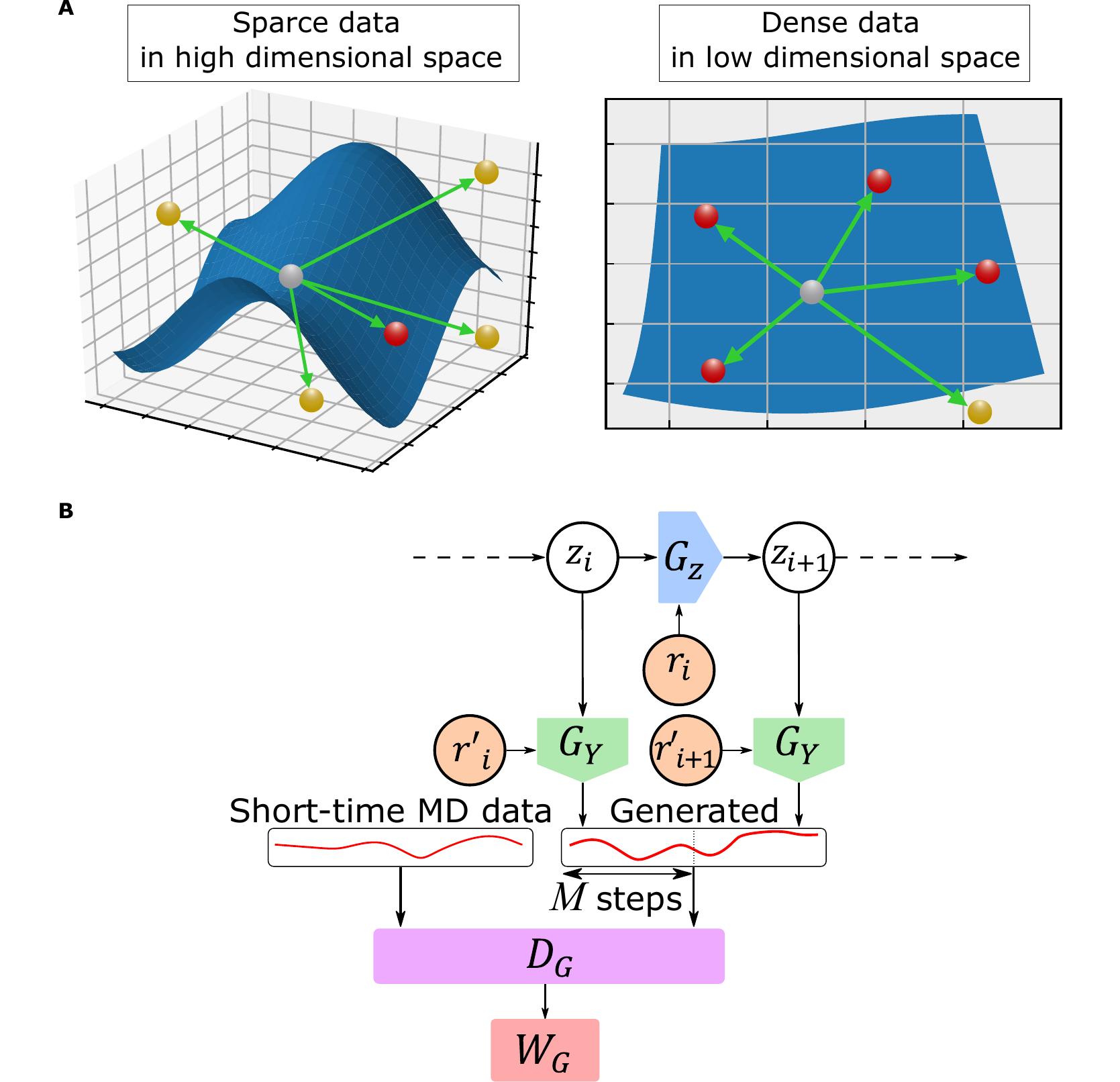}
\caption{
{\bf Architecture of MD-GAN.}
{\bf (A)} Schematic view of time evolution in a low-dimensional space. The gray, red, and yellow spheres contain the trajectory information for $M$ steps of the extracted subsystem. The blue region is the manifold where the information of $M$ steps can be found. When the gray sphere has information of the current $M$ steps, it can take various states via stochastic transition, during the next time evolution. In this evolution, although the red sphere is on the manifold, the yellow sphere is off of it. Thus, taking the yellow sphere leads to an increase in bias. This bias can be mitigated by time evolving in a low-dimensional space.
{\bf (B)} Structure of MD-GAN. $z_i$ is a latent variable, $r_i$ and $r'_i$ are random numbers, $G_z$ is the generator for time evolution in latent space, and $G_Y$ is the generator for generating trajectories from latent variables. $W_G$ is the Wasserstein distance calculated by the discriminator $D_G$.
}
\label{fig2}
\end{center}
\end{figure*}

\begin{figure*}[t!]
\begin{center}
\includegraphics[width=7 cm]{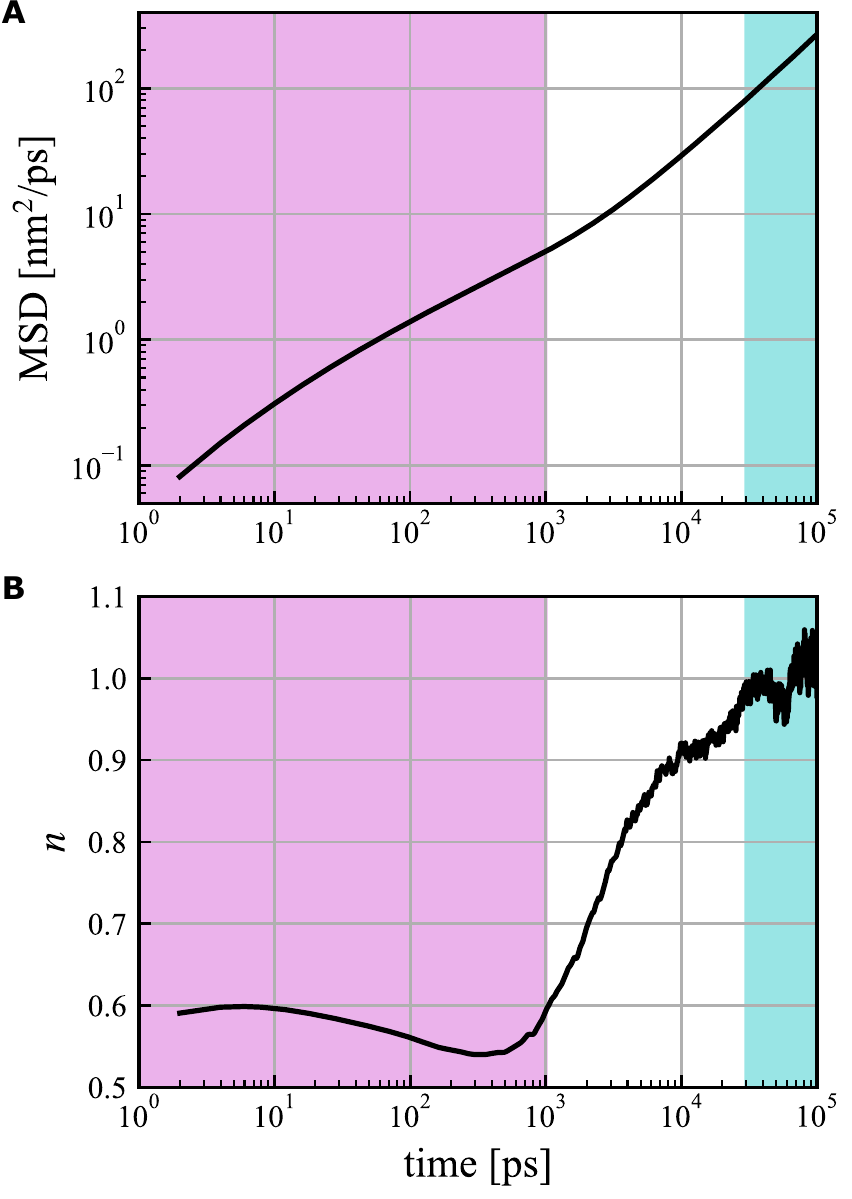}
\caption{
{\bf Mean Square Displacement (MSD) of polyethylene.}
{\bf (A)} MSD directory estimated from MD simulation data. The pink region represents the time range where the MSD is proportional to the square root of $t$, and the light blue region represents the time range where the MSD is linear (normal diffusion).
{\bf (B)} Relationship MSD \& $t$ is $MSD(t)\propto t^n$, assuming MSD is proportional to $t$ to the $n$th power.
}

\label{fig3}
\end{center}
\end{figure*}

\begin{figure*}[t!]
\begin{center}
\includegraphics[width=7 cm]{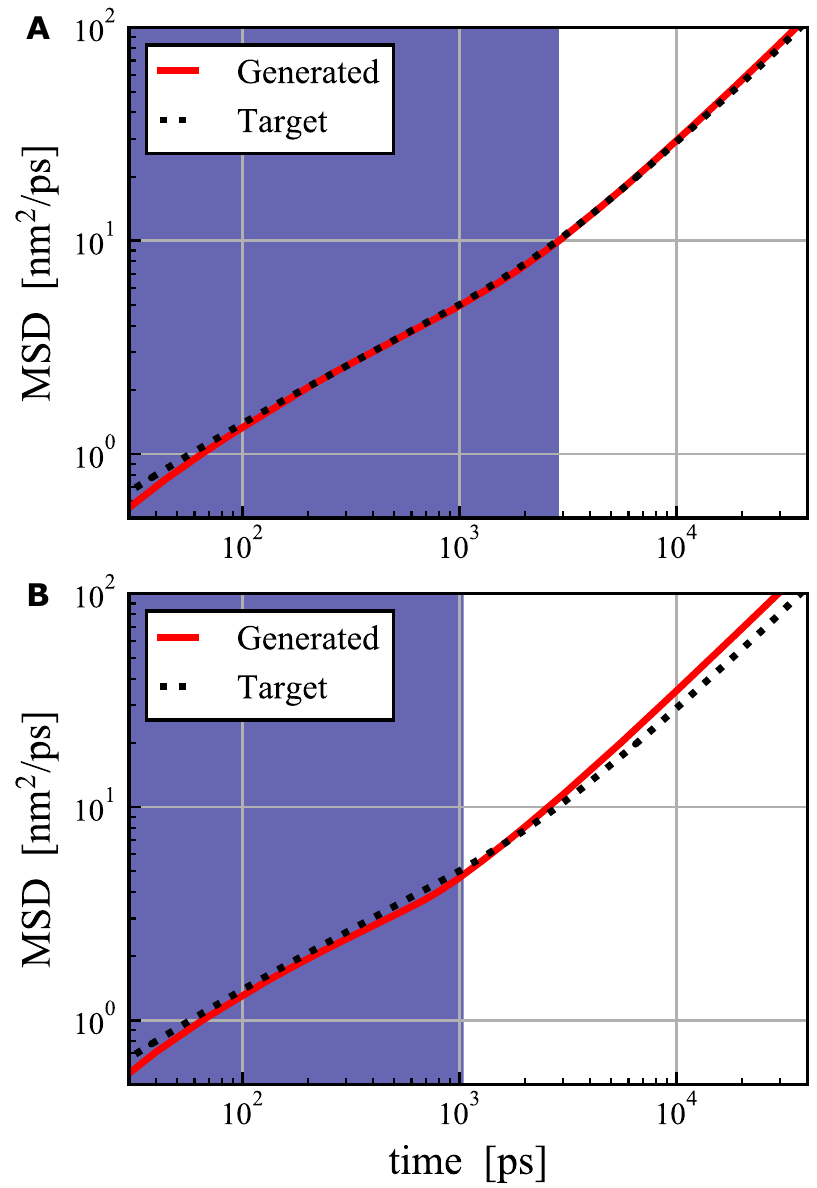}
\caption{
{\bf The MSD obtained by MD-GAN with single particle input.}
The red solid line represents the MSD estimated from the trajectories generated by MD-GAN, and the black dashed line represents the same estimated from the trajectories of long-time MD. The purple area corresponds to the length of the short-time MD data input to MD-GAN.
{\bf (A)} MSD obtained by the prediction using MD data for 2816 ps. The $err$ for this MSD was 1.6 \%.
{\bf (B)} MSD obtained by the prediction using MD data for 1024 ps. The $err$ for this MSD was 14.7 \%.
}
\label{fig4}
\end{center}
\end{figure*}

\begin{figure*}[!t]
\begin{center}
\includegraphics[width=\linewidth]{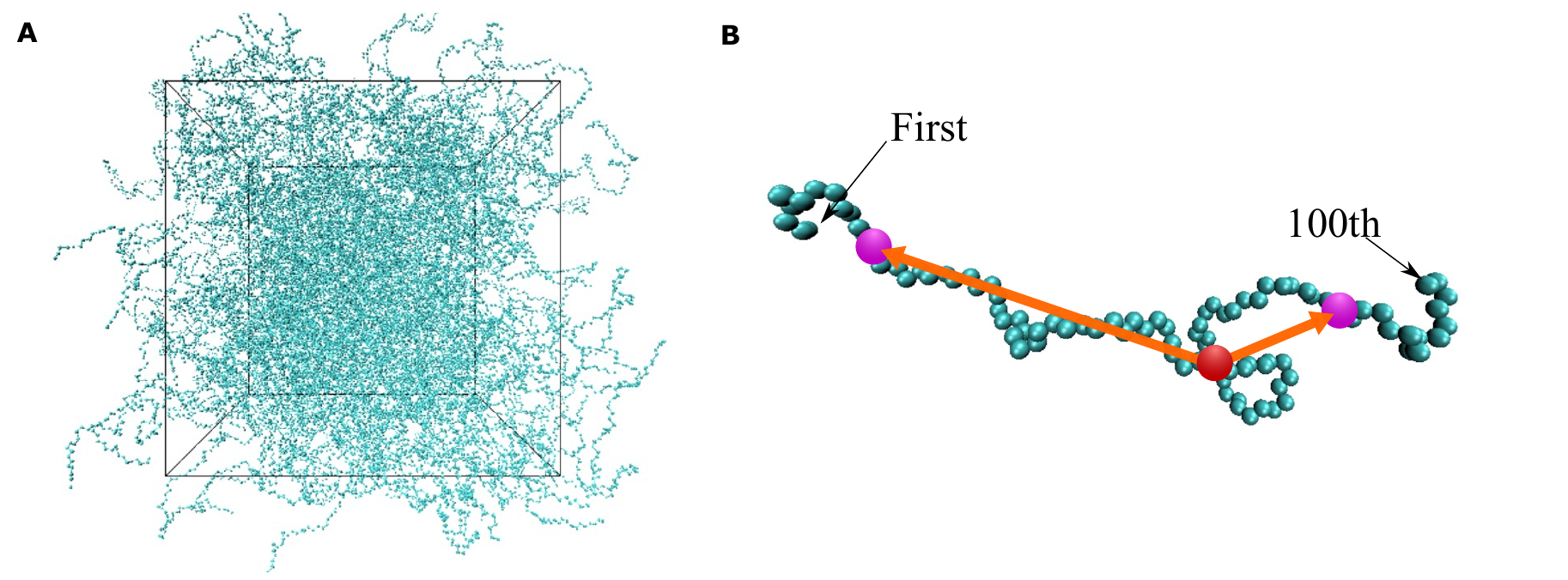}
\caption{
{\bf Three-particle input for the PE system.}
{\bf (A)} Snapshot of the system.
{\bf (B)} The three extracted particles and the input information. The red particle represents the center of the chain, and the two pink particles represent the 15th and 85th particles. The orange vectors provide the coordinates of the pink particles relative to the red particle. For the three-particle input, the coordinates of the center of the chain and the two relative coordinates were used as input.
}
\label{fig5}
\end{center}
\end{figure*}

\begin{figure*}[!t]
\begin{center}
\includegraphics[width=7 cm]{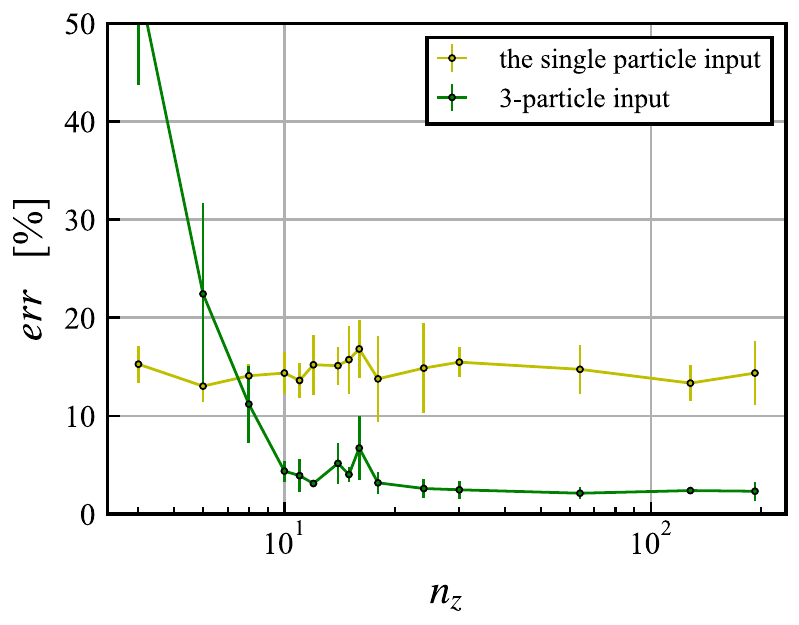}
\caption{
{\bf Relationship between the dimensions of latent variables $n_z$ and $err$.} $err$ is calculated from the prediction results using MD data for 1024 ps as input. The yellow and green lines represent the minimum $err$ for the single particle input and the three-particle input, respectively.
}
\label{fig6}
\end{center}
\end{figure*}

\begin{figure*}[!t]
\begin{center}
\includegraphics[width=7 cm]{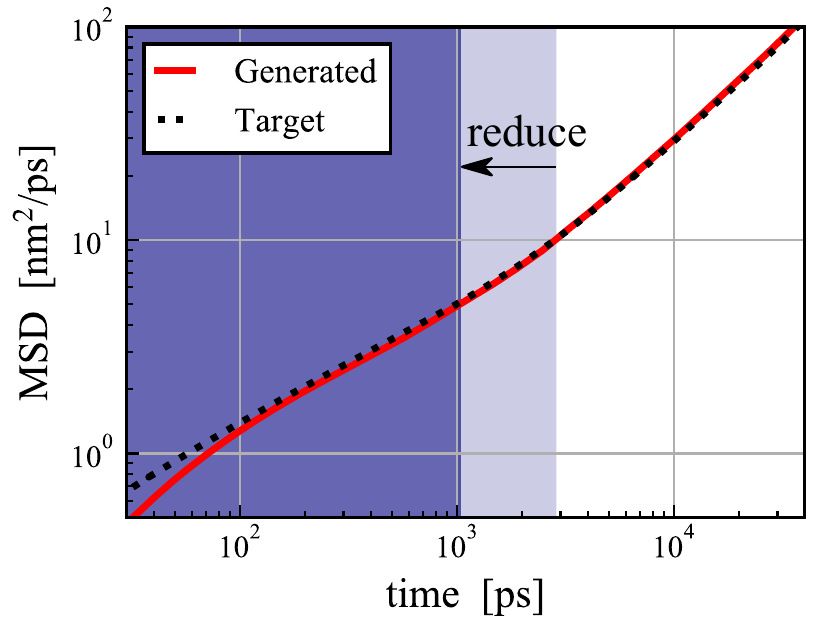}
\caption{
{\bf The MSD obtained by MD-GAN with three-particle input.}
The MSD obtained by MD-GAN with three-particle input. The dark purple area corresponds to the time length of the input data for three-particle input (1024 ps). The light purple region corresponds to the reduced length of the short-time MD data due to the change from single-particle to multi-particle input. The $err$ for this MSD was 2.0 \%. As shown in Figure 4B, 2816 ps of MD data was required for accurate prediction using single particle input. Thus, the time length of input data was reduced to approximately one third (light purple region).
}
\label{fig7}
\end{center}
\end{figure*}

\end{document}